\documentclass[showkeys,showpacs,prd,aps,floats,eqsecnum,preprint]{revtex4}

\begin{document}

\title{On the energy transported by exact plane\\ gravitational-wave
solutions}

\author{Yuri N.~Obukhov}
\email{yo@ift.unesp.br}
\affiliation{Instituto de F\'{\i}sica Te\'orica,
UNESP-Universidade Estadual Paulista,
Rua Dr.\ Bento Teobaldo Ferraz 271, 01140-070 S\~ao Paulo, Brazil}
\affiliation{Department of Theoretical Physics,
Moscow State University, 117234
Moscow, Russia}

\author{J.~G.~Pereira}
\email{jpereira@ift.unesp.br}
\affiliation{Instituto de F\'{\i}sica Te\'orica,
UNESP-Universidade Estadual Paulista,
Rua Dr.\ Bento Teobaldo Ferraz 271, 01140-070 S\~ao Paulo, Brazil}

\author{Guillermo F.~Rubilar}
\email{grubilar@udec.cl}
\affiliation{Departamento de F{\'{\i}}sica,
Universidad de Concepci\'on, Casilla 160-C, Concepci\'on, Chile}

\begin{abstract}
The energy and momentum transported by {\em exact} plane
gravitational-wave solutions of Einstein equations are computed
using the teleparallel equivalent formulation of Einstein's theory.
It is shown that these waves transport neither energy nor momentum.
A comparison with the usual {\em linear} plane gravitational-waves
solution of the linearized Einstein equation is presented.
\end{abstract}


\maketitle

%%%%%%%%%%%%%%%%%%%%%%%
\section{Introduction}
%%%%%%%%%%%%%%%%%%%%%%%

Although there are compelling indirect experimental evidences \cite{HT} of
the existence of gravitational waves \cite{maggiore}, they have not yet 
been
directly detected. These evidences, it is important to remark, do not give
a clue on the form and properties of the gravitational waves. Since the
potential astrophysical sources of gravitational waves are at enormous 
distances
{}from us, the amplitude of a wave when reaching a detector on Earth 
should be so
small that the linearized theory is usually assumed to provide a 
sufficiently
accurate description of these waves. In other words, to comply with the 
idea
that the radiative solutions satisfy a linear wave equation, one should 
assume
that these waves do not carry enough energy and momentum to affect their 
own
plane-wave form \cite{weinberg}.

The search for gravitational waves relies strongly on the assumptions 
above. The
lack of direct observational evidences, however, may question these 
assumptions.
In particular, it is well known that a field can transport its own 
source only
via an essentially nonlinear process \cite{bondi}. This point is the 
origin of
a long-standing discussion on the energy and momentum transported by
gravitational waves \cite{cs1}.

The study of gravitational waves in general relativity has a long and rich
history \cite{peres,pen,griff,vdz}. The discussion of the
possible generalizations of such solutions revealed the exact wave 
solutions
in Poincar\'e gauge gravity \cite{adam,chen,sippel,vadim,singh,babu},
in teleparallel gravity \cite{tele}, in generalized Einstein theories
\cite{gurses2,lovelock}, in metric-affine theory of gravity
\cite{dirk1,dirk2,king,pasic,vas1,vas2,ppMAG}, in supergravity
\cite{sg1,sg2,sg3,sg4,sg5}, as well as, more recently, in superstring
theories \cite{gimon,ark,str}. Some attention has
also been paid to the higher-dimensional generalizations of the
gravitational wave solutions \cite{coley,hervik,ppN}.

On the other hand, plane-fronted gravitational waves represent an important
class of exact solutions of Einstein's equation \cite{exact1,exact2,exact}.
In this paper we consider the question of the energy-momentum transported
by such plane-wave solutions. Of course, any exact gravitational wave 
solution
is highly idealized, and is thus probably more of academic interest 
\cite{mtw}.
Nevertheless, we expect that our study may shed some new light on that 
question,
as well as to develop a further understanding of the nonlinear regime of 
gravitation.
Our general notations are as in \cite{HMMN95}. In particular, we use the
Latin indices $i,j,\dots$ for local holonomic spacetime coordinates and the
Greek indices $\alpha,\beta,\dots$ label (co)frame components. Particular
frame components are denoted by hats, $\hat 0, \hat 1$, etc. As usual,
the exterior product is denoted by $\wedge$, while the interior product 
of a
vector $\xi$ and a $p$-form $\Psi$ is denoted by $\xi\rfloor\Psi$. The 
vector
basis dual to the frame 1-forms $\vartheta^\alpha$ is denoted by $e_\alpha$
and they satisfy $e_\alpha\rfloor\vartheta^\beta=\delta^\beta_\alpha$.
Using local coordinates $x^i$, we have $\vartheta^\alpha=h^\alpha_idx^i$ 
and
$e_\alpha=h^i_\alpha\partial_i$. We define the volume 4-form by
$\eta:=\vartheta^{\hat{0}}\wedge\vartheta^{\hat{1}}\wedge
\vartheta^{\hat{2}}\wedge\vartheta^{\hat{3}}$. Furthermore, with the 
help of
the interior product we define $\eta_{\alpha}:=e_\alpha\rfloor\eta$,
$\eta_{\alpha\beta}:=e_\beta\rfloor\eta_\alpha$, $\eta_{\alpha\beta\gamma}:=
e_\gamma\rfloor\eta_{\alpha\beta}$ which are bases for 3-, 2- and 1-forms
respectively. Finally, $\eta_{\alpha\beta\mu\nu} = e_\nu\rfloor\eta_{\alpha
\beta\mu}$ is the Levi-Civita tensor density. The $\eta$-forms satisfy the
useful identities:
\begin{eqnarray}
\vartheta^\beta\wedge\eta_\alpha &=& \delta_\alpha^\beta\eta ,\\
\vartheta^\beta\wedge\eta_{\mu\nu} &=& \delta^\beta_\nu\eta_{\mu} -
\delta^\beta_\mu\eta_{\nu},\label{veta1}\\ \label{veta}
\vartheta^\beta\wedge\eta_{\alpha\mu\nu}&=&\delta^\beta_\alpha\eta_{\mu\nu
} + \delta^\beta_\mu\eta_{\nu\alpha} + \delta^\beta_\nu\eta_{\alpha\mu},\\
\label{veta2}
\vartheta^\beta\wedge\eta_{\alpha\gamma\mu\nu}&=&\delta^\beta_\nu\eta_{\alpha
\gamma\mu} - \delta^\beta_\mu\eta_{\alpha\gamma\nu} + \delta^\beta_\gamma
\eta_{\alpha\mu\nu} - \delta^\beta_\alpha\eta_{\gamma\mu\nu}.
  \end{eqnarray}
The line element $ds^2 =
g_{\alpha\beta}\vartheta^\alpha\otimes\vartheta^\beta$ is defined by the
spacetime metric $g_{\alpha\beta}$ of signature $(+,-,-,-)$.

%%%%%%%%%%%%%%%%%%%%%%%
\section{Teleparallel gravity}
%%%%%%%%%%%%%%%%%%%%%%%

The teleparallel approach is based on the gauge theory of
translations, and the coframe $\vartheta^\alpha = h^\alpha_idx^i$
(tetrad) plays the role of the corresponding gravitational
potential. Einstein's general relativity theory can be reformulated
as the teleparallel theory. Geometrically, one can view the
teleparallel gravity as a special (degenerate) case of the
metric-affine gravity in which the coframe $\vartheta^\alpha$ and
the local Lorentz connection $\Gamma_\alpha{}^\beta$ are subject
to the distant parallelism constraint $R_\alpha{}^\beta = 0$ 
\cite{telemag}.
The torsion 2-form
\begin{equation}
T^\alpha = d\vartheta^\alpha+\Gamma_\beta{}^{\alpha}\wedge\vartheta^\beta,
\label{defT}
\end{equation}
arises as the gravitational gauge field strength.
In the so-called teleparallel equivalent gravity model, the Lagrangian 
reads:
\begin{equation} \label{V2}
V = -\,{\frac 1 {2\kappa}}T^{\alpha}\wedge{}^\star\left(T_{\alpha}
- \vartheta^\alpha\wedge e_\beta\rfloor T^\beta -{1\over 2}\,e^\alpha\rfloor
(\vartheta^\beta\wedge T_\beta)\right).
\end{equation}
Here $\kappa=8\pi G/c^3$, and ${}^\star$ denotes the Hodge duality operator
in the metric $g_{\alpha\beta}$. The latter is assumed to be the flat 
Minkowski
metric $g_{\alpha\beta} = o_{\alpha\beta} :={\rm diag}(+1,-1,-1,-1)$, and we
use it to raise and lower local frame (Greek) indices.

The teleparallel field equations are obtained from the variation of
the total action with respect to the coframe,
\begin{equation}
DH_\alpha - E_\alpha = \Sigma_\alpha,\label{geq}
\end{equation}
where $DH_\alpha=d H_\alpha-\Gamma_\alpha^{\ \beta}\wedge H_\beta$ denotes
the covariant exterior derivative. The translational momentum
and the canonical energy-momentum are, respectively:
\begin{eqnarray}
H_{\alpha} = -\,{\frac {\partial V} {\partial T^\alpha}} &=&
\,{1\over \kappa}\,{}^\star\!\left(T_{\alpha} - \vartheta_\alpha\wedge
e_\beta\rfloor T^\beta -{1\over 2}\,e_\alpha\rfloor
(\vartheta^\beta\wedge T_\beta)\right),\label{Ha0}\\
E_\alpha = {\frac {\partial V} {\partial \vartheta^\alpha}} &=&
e_\alpha\rfloor V + (e_\alpha\rfloor T^\beta)\wedge H_\beta.\label{Ea0}
\end{eqnarray}
The Lagrangian (\ref{V2}) can then be recast as
\begin{equation}
V = -\,{\frac 12}\,T^\alpha\wedge H_\alpha.\label{Vlag}
\end{equation}
It should be mentioned that the resulting model is degenerate, from the
metric-affine viewpoint,
because the variational derivatives of the action with the respect to
the metric and connection are trivial. This means that the field
equations are satisfied for any $\Gamma_\alpha{}^\beta$. However, the
presence of the connection field plays an important role that may be
characterized as a {\it regularization}. The latter is twofold.

First of all, teleparallel gravity becomes explicitly covariant
under the local Lorentz transformations of the coframe. In particular,
the Lagrangian (\ref{V2}) is invariant under the change of variables
\begin{equation}\label{cofcontrans}
\vartheta'^\alpha = L^\alpha{}_{\beta}\vartheta^\beta,\qquad
\Gamma'_\alpha{}^{\beta} = (L^{-1})^\mu{}_{\alpha}\Gamma_\mu{}^\nu
L^\beta{}_{\nu} + L^\beta{}_{\gamma}d(L^{-1})^\gamma{}_{\alpha},
\end{equation}
with $L^\alpha{}_{\beta}(x)\in SO(1,3)$. In contrast, for the pure tetrad
gravity, which is obtained when we put $\Gamma_\alpha{}^\beta = 0$,
the Lagrangian is only quasi-invariant---it changes by a total divergence.
The connection 1-form $\Gamma_\alpha{}^\beta$ can be decomposed into the
Riemannian and post-Riemannian parts as
\begin{equation}
\Gamma_\alpha{}^\beta = \tilde{\Gamma}_\alpha{}^\beta -
K_\alpha{}^\beta .\label{gagaK}
\end{equation}
Here $\tilde{\Gamma}_\alpha{}^\beta$ is the purely Riemannian connection
and $K_\alpha{}^\beta$ is the contortion 1-form which is related to the
torsion through the identity
\begin{equation}
T^\alpha=K^\alpha{}_\beta\wedge\vartheta^\beta.\label{contor}
\end{equation}
Then one can show that due to geometric identities \cite{PGrev}, the gauge
momentum (\ref{Ha0}) can be written as
\begin{equation}
H_\alpha ={\frac 1 
{2\kappa}}\,K^{\mu\nu}\wedge\eta_{\alpha\mu\nu}.\label{H0K}
\end{equation}

A second, and even more important property of the teleparallel framework is
that the Weitzenb\"ock connection actually represents inertial
effects that arise due to the choice of the reference system \cite{apl}. 
The
inertial contributions in many cases yield unphysical results for the
total energy of the system, producing either trivial or divergent
answers. The teleparallel connection acts
as a regularizing tool which helps to subtract the inertial effects
without distorting the true gravitational contribution \cite{rtg}.

%%%%%%%%%%%%%%%%%%%%%%%
\section{Energy-momentum conservation}
%%%%%%%%%%%%%%%%%%%%%%%

We begin by rewriting the field equation (\ref{geq}) in the Maxwell-type
form:
\begin{equation}
DH_\alpha = E_\alpha + \Sigma_\alpha.\label{DH0}
\end{equation}
{}The analogy with the electromagnetism is obvious. The Maxwell 2-form
$F=dA$ represents the gauge field strength of the electromagnetic
potential 1-form $A$. From the Lagrangian $V(F)$, the 2-form of the
electromagnetic excitations is defined by $H = - \partial V/\partial
F$, and the field equation reads $dH = J$, where $J$ is the 3-form of
the electric current density of matter. In view of the nilpotency of
the exterior differential, $dd \equiv 0$, the Maxwell equation yields
the conservation law of the electric current, $dJ = 0$.

In contrast to electrodynamics, gravity is fundamentally nonlinear.
It is more akin to a Yang-Mills theory, though the ``internal'' index of
the gauge field potential 1-form $\vartheta^\alpha$ is not really
internal, but of spacetime nature.
The gauge field strength 2-form $T^\alpha = D\vartheta^\alpha$
is now defined by the covariant derivative of the potential (compare
with $F=dA$). The gravitational field excitation 2-form $H_\alpha$ is
introduced by (\ref{Ha0}), in a direct analogy to the Maxwell theory
(recall $H = - \partial V/\partial F$). Finally, we observe that as
compared to the Maxwell field equation $dH = J$, the gravitational
field equation (\ref{DH0}) contains now the covariant derivative $D$,
and in addition, the right-hand side is represented by a modified current
3-form, $E_\alpha + \Sigma_\alpha$. The last term is the energy-momentum
of matter, and we naturally conclude that the 3-form $E_\alpha$ describes
the {\it energy-momentum current of the gravitational field}. Its
presence in the right-hand side of the field equation (\ref{DH0})
reflects the self-interacting nature of the gravitational field.

We can complete the analogy with electrodynamics by deriving the
corresponding conservation law. Indeed, since $DD \equiv 0$ for the
teleparallel connection, (\ref{DH0}) tells us that the sum of the
energy-momentum currents of gravity and matter, $E_\alpha + \Sigma_\alpha$,
is covariantly conserved \cite{apl},
\begin{equation}
D(E_\alpha + \Sigma_\alpha) = 0.\label{DE1}
\end{equation}
This covariant conservation law is consistent with the covariant
transformation properties of the currents $E_\alpha$ and $\Sigma_\alpha$.

One can rewrite the conservation of energy-momentum in terms of the
ordinary derivatives. Using the explicit expression $DH_\alpha = dH_\alpha
- \Gamma_\alpha{}^\beta\wedge H_\beta$, the field equation (\ref{geq})
and (\ref{DH0}) can be recast in an alternative form
\begin{equation}
dH_\alpha = {\cal E}_\alpha + \Sigma_\alpha,\label{DH1}
\end{equation}
where ${\cal E}_\alpha = E_\alpha + \Gamma_\alpha{}^\beta\wedge H_\beta$.
Accordingly, Eq.~(\ref{DH1}) yields a usual conservation law with the
ordinary derivative,
\begin{equation}
d({\cal E}_\alpha + \Sigma_\alpha) = 0.\label{DE2}
\end{equation}

The 3-form $E_\alpha$ describes the gravitational energy-momentum in a
covariant way, whereas the 3-form ${\cal E}_\alpha$ is a non-covariant
object. In terms of components, ${\cal E}_\alpha$ gives rise to the energy-
momentum
pseudotensor. It is worthwhile to note that $H_\alpha$ thus plays a
role of the energy-momentum superpotential both for the total
covariant energy-momentum current $(E_\alpha + \Sigma_\alpha)$ and
for the noncovariant current $({\cal E}_\alpha + \Sigma_\alpha)$.

The $\eta$-forms (defined above) serve as the basis of the spaces of forms
of different rank, and when we expand the above objects with respect to
the $\eta$-forms, the usual tensor formulation is recovered. Explicitly,
\begin{equation}
H_\alpha = {\frac 1\kappa}\,S_\alpha{}^{\mu\nu}\,\eta_{\mu\nu}.\label{Ha}
\end{equation}
Here $S_\alpha{}^{\mu\nu} = - S_\alpha{}^{\nu\mu}$ is constructed from the
contortion tensor in the usual way.

Analogously, we have explicitly for the gravitational energy-momentum
\begin{equation}
E_\alpha = {\frac 12}\left[(e_\alpha\rfloor T^\beta)\wedge H_\beta
- T^\beta\wedge(e_\alpha\rfloor H_\beta)\right].\label{Ea00}
\end{equation}
Substituting here (\ref{Ha}) and $T^\alpha = {\frac 12}\,T_{\mu\nu}{}^\alpha
\,\vartheta^\mu\wedge\vartheta^\nu$ and using 
Eqs.~(\ref{veta1})-(\ref{veta2}),
we find \cite{apl}
\begin{equation}
E_\alpha = t_\alpha{}^\beta\,\eta_\beta,\qquad t_\alpha{}^\beta =
{\frac 1{2\kappa}}\left(4T_{\alpha\nu}{}^\lambda S_\lambda{}^{\beta\nu}
- T_{\mu\nu}{}^\lambda S_\lambda{}^{\mu\nu}\,\delta_\alpha^\beta
\right).\label{Ea1}
\end{equation}
Similarly, we find
\begin{equation}
{\cal E}_\alpha = j_\alpha{}^\beta\,\eta_\beta,\qquad j_\alpha{}^\beta =
{\frac 1{2\kappa}}\left(4T_{\alpha\nu}{}^\lambda S_\lambda{}^{\beta\nu}
- T_{\mu\nu}{}^\lambda S_\lambda{}^{\mu\nu}\,\delta_\alpha^\beta
+ 4\Gamma_{\nu\alpha}{}^\lambda\,S_\lambda{}^{\beta\nu}\right).\label{Ea2}
\end{equation}
Taking into account the analogous expansion of the matter energy-momentum,
$\Sigma_\alpha = \Sigma_\alpha{}^\beta\,\eta_\beta$, that introduces the 
tensor
of energy-momentum, and using (\ref{Ha}) and (\ref{Ea1}), we easily recover
the field equation in tensor language (used, for example, in \cite{dAGP00}).
Note that the conservation laws (\ref{DE1}) and (\ref{DE2}) coincide when
we put $\Gamma_\alpha{}^\beta = 0$. The last term in (\ref{Ea2}) then
disappears, whereas the torsion reduces to the anholonomity 2-form,
$T^\alpha = F^\alpha = d\vartheta^\alpha$ \cite{conserved,noninv}.

%%%%%%%%%%%%%%%%%%%%%%%
\section{Exact plane wave}
%%%%%%%%%%%%%%%%%%%%%%%

In the local coordinates $(\sigma,\rho,y,z)$, the plane-fronted 
gravitational
wave \cite{peres,pen} is described by the coframe components:
\begin{equation}
\vartheta^{\widehat{0}} = (1 + h/4)d\sigma + d\rho,\qquad
\vartheta^{\widehat{1}} = (1 - h/4)d\sigma - d\rho,\qquad
\vartheta^{\widehat{2}} = dy,\qquad
\vartheta^{\widehat{3}} = dz.\label{cof}
\end{equation}
When the function $h(\sigma,y,z)$ satisfies the two-dimensional
Laplace equation,
\begin{equation}
{\frac {\partial^2 h}{\partial y^2}}
+ {\frac {\partial^2 h}{\partial z^2}} = 0,
\end{equation}
the coframe (\ref{cof}) represents an exact solution of Einstein's
field equation in vacuum. In particular, one usually chooses
$h = yz\,u(\sigma)$ or $h = (y^2 - z^2)v(\sigma)$, where $u,v$ are
arbitrary functions of $\sigma$.

This configuration describes a plane-fronted gravitational wave that
is characterized by the wave covector $k = k_\alpha\vartheta^\alpha
= 2d\sigma$, so that the wave fronts are represented by the plane
surfaces of constant $\sigma$. The coordinate $\rho$ is an affine
parameter along the wave vector of the null geodesic. As we see,
the tetrad components of the wave covector read $k_\alpha = (1,1,0,0)$,
and hence $k^\alpha = (1,-1,0,0)$. Obviously, $k^\alpha k_\alpha = 0$.

%%%%%%%%%%%%%%%%%%%%%%%
\section{Energy transported by an exact plane wave}
%%%%%%%%%%%%%%%%%%%%%%%

The coframe (\ref{cof}) becomes holonomic when gravitation is
switched off. In technical terms this means that its anholonomy
is not related to inertial effects, but only to gravitation.
Accordingly, we conclude that in this frame the Weitzenb\"ock
connection can be consistently put equal to zero:
$\Gamma_\alpha{}^\beta =0$. Torsion then reads
\begin{equation}
T^\alpha = {\frac {k^\alpha}4}\,d\sigma\wedge dh.
\end{equation}
In this case, Eq.~(\ref{Ha}) yields
\begin{equation}
H_\alpha = -\,{\frac {k_\alpha}4}\,d\sigma\wedge{}^{\underline{\star}}dh,
\end{equation}
where ${}^{\underline{\star}}$
denotes the Hodge operator in the 2-dimensional Euclidean plane
$(y,z)$, namely, ${}^{\underline{\star}}(a \, dy + b \,dz) = a \, dz - b 
\, dy$.
Hence, from Eq.~(\ref{Ea00}) we see that
\begin{equation}
E_\alpha = 0.
\end{equation}
This is a covariant result that does not depend on the choice of the
reference system. When we choose a different reference system, the
corresponding coframe is related by a Lorentz transformation to the
original one, and the teleparallel connection in this system is
computed from (\ref{cofcontrans}). It will be nontrivial, in general,
thus reflecting the possible non-inertiality of the reference system.
Since under the Lorentz transformation (\ref{cofcontrans}) the 3-form
of the gravitational energy-momentum transforms covariantly,
$E'_\alpha = (L^{-1})^\beta{}_\alpha E_\beta$, it will be zero in all
reference systems.

Earlier, it was demonstrated \cite{gurses} that it is possible to
choose local coordinates in such a way that the pseudotensor of
the energy-momentum of plane gravitational waves vanishes. Our result
is, however, much stronger. Due to the use of the exterior calculus,
all our computations are independent of the choice of the local
coordinates. In addition, the covariance of the framework under
the local Lorentz transformations makes our results independent
also of the choice of the reference system. We can then conclude that
the exact plane-wave solution (\ref{cof}) transports neither energy nor
momentum.

%%%%%%%%%%%%%%%%%%%%%%%
\section{Remark on superenergy}
%%%%%%%%%%%%%%%%%%%%%%%

In the studies of gravitational waves, there is a long history of the
attempts to describe the energy and momentum of the pure radiation by the
so called superenergy tensors; the most well-known of them is perhaps the
Bel-Robinson tensor \cite{BR}. In the case of general relativity, the 
corresponding
conserved current 3-form is defined by
\begin{equation}
B_{\alpha\mu\nu} = {\frac 12}\left[(e_\alpha\rfloor
\widetilde{R}{}_\mu{}^\lambda)\wedge{}^\star\widetilde{R}_{\nu\lambda} 
- \widetilde{R}{}_\nu{}^\lambda\wedge(e_\alpha
\rfloor{}^\star\widetilde{R}_{\mu\lambda})\right], \label{BRform}
\end{equation}
where $\widetilde{R}{}_\mu{}^\nu = {\frac 
12}\,\widetilde{R}{}_{\alpha\beta\mu}{}^\nu
\,\vartheta^\alpha\wedge\vartheta^\beta$ is the Riemannian curvature 2-form.
There is an obvious similarity between (\ref{BRform}) and the definition 
of the
gravitational energy-momentum current (\ref{Ea00}). It is also similar 
to the
energy-momentum tensor of the
electromagnetic field,
\begin{equation}
E^{\rm (em)}_\alpha = {\frac 12}\left[(e_\alpha\rfloor F)
\wedge H - F\wedge (e_\alpha\rfloor H)\right],
\end{equation}
with $F$ the Maxwell field
strength 2-form, and $H = {}^\star F$ the electromagnetic excitation 2-form.
Notice, however, that the superenergy carries more indices than the 
energy-momentum
forms.

Using the explicit form (\ref{cof}) of the plane-wave metric in the 
Bel-Robinson tensor
(\ref{BRform}), we find
\begin{equation}\label{Bamn}
B_{\alpha\mu\nu} = |\gamma|^2\,k_\alpha k_\beta k_\mu k_\nu\,\eta^\beta.
\end{equation}
Here $|\gamma|^2 := {}^\star(\gamma_\alpha\wedge{}^\star\gamma^\alpha)$,
where in accordance with Ref.~\cite{ppN} we introduce the 1-form 
$\gamma_\alpha$ by $\gamma_{\hat 0} = \gamma_{\hat 1} = 0$ and $\gamma_a 
= - {\frac 12}d(e_a\rfloor dh)$, for $a = 2,3$. This 1-form is the key
element that determines the structure of the Riemannian curvature of the
plane-wave solution (see the eq. (7) of \cite{ppN}), and it describes
the polarization properties of the wave. The factor $|\gamma|^2$ is a 
positive quantity, and for the specific choices of the function $h$ 
mentioned above, $|\gamma|^2$ is proportional to $u^2(\sigma)$ and
$v^2(\sigma)$. Thus indeed the Bel-Robinson form has a nontrivial value 
for the plane waves, with the structure of (\ref{Bamn}) resembling the 
expressions for the energy of the electromagnetic waves.

Unfortunately, however, the dynamical role of the Bel-Robinson 
superenergy, despite a
number of nice properties, is still unclear. In particular, the main problem
remains on the choice of the overall factor in (\ref{BRform}), which is 
needed to
obtain the correct dimension. In this respect, it is worthwhile to mention a
recent paper \cite{DTbel} that addresses the problem of finding a dynamical
framework for the Bel-Robinson superenergy.

%%%%%%%%%%%%%%%%%%%%%%%
\section{Discussion and conclusion}
%%%%%%%%%%%%%%%%%%%%%%%

As we have seen, the exact plane-wave solution of Einstein's equation
transports neither energy nor momentum. This property appears to be quite
unusual, since one could expect that a gravitational wave, like any other
gravitational field configuration, should be characterized by nontrivial
distribution of energy and momentum. In this sense the exact plane wave
solution, although being mathematically well defined, appears to be a
physical puzzle. A detailed discussion of its meaning may thus be needed.
At the moment it is unclear whether such a property is shared by other
exact wave solutions known in the literature, although the fact that the 
plane wave configuration is simultaneously a solution of full Einstein's
equation and of the linearized equation appears to be an important property
that most probably yields the vanishing of the energy. 

One can wonder whether the inability to transport
energy and momentum is a property of this specific exact solution, or it
is a general property of any linear gravitational waves. In order to get
some insight into this question, a comparison with electromagnetism is
quite elucidative. Let us rewrite Maxwell equation as follows
\begin{equation}
d H = E + J,
\end{equation}
where we introduced the electric self-interaction current $E$ along with 
the
usual electric matter current 3-form $J$. As is well known, 
electromagnetism
is linear, and since any self-current is at least quadratic in the field
variables, the electromagnetic self-interaction term vanishes identically:
$E=0$. Due to this linearity, an electromagnetic wave does not transport 
its
own source, that is, the electric charge. Of course, electromagnetic waves
are able to transport energy and momentum, since neither energy nor 
momentum
is a source of the electromagnetic field, and the energy-momentum current
does not appear in the electromagnetic field equation. Accordingly, the
linearity of electromagnetism does not restrict the energy-momentum current
to be linear. In fact, the energy transported by an electromagnetic wave
is given by the quadratic Poynting vector.

The picture changes significantly for gravitation. The crucial difference
is that energy-momentum is source of gravitation, and consequently
{\it the energy-momentum self-interaction current $E_\alpha$ appears 
explicitly
in the gravitational field equations}
\begin{equation}
DH_\alpha = E_\alpha + \Sigma_\alpha.
\label{DH0Bis}
\end{equation}
In the linear approximation, the energy-momentum current is restricted to
be linear, and consequently it vanishes in that approximation \cite{gw1}.
We can then say that, similarly to the fact that electromagnetic waves are
unable to transport its own (electric) charge, a linear gravitational wave
seems to be unable to transport its ``charge", that is, the 
energy-momentum.
Usually one takes the second-order energy-momentum current, but this is a
questionable procedure because such a current appears as source only in the
second-order wave equation. It should, by this reason, represent the
energy-momentum transported by the second-order nonlinear gravitational 
wave.

Our results do not mean at all that gravitational waves cannot exist.
However, they seem to support the idea that the true gravitational wave
should be essentially a nonlinear physical phenomenon. The analysis of the
specific example of the exact plane wave considered here shows that plane
waves seem to have rather unusual properties, such as the vanishing of 
their
energy-momentum current \cite{FaPa}. In the light of our observations, 
their
physical meaning remains uncertain. Finally, we would like to add that, 
while
technically obtained in the context of the teleparallel equivalent 
formulation
of Einstein's theory, this result is physically meaningful, and as such 
it might
also be true in the context of general relativity.

%%%%%%%%%%%%%%%%%%%%%%%%%%
\begin{acknowledgments}
The work of YNO was supported by a grant from FAPESP. The authors thank
FAPESP, CNPq and CAPES for partial financial support.
\end{acknowledgments}

\end{document}